\documentclass[letterpaper]{article}%
\usepackage{amsmath}%
\usepackage{amsfonts}%
\usepackage{amssymb}%
\usepackage{graphicx}
\usepackage[draft]{hyperref}
\usepackage[margin=2.5cm]{geometry}
\usepackage[rgb]{xcolor}

\begin{document}

\title{\textbf{Optical fiber tips functionalized with semiconductor photonic crystal cavities}}
\author{Gary Shambat,\footnote{email: gshambat@stanford.edu}\hspace{1.5 mm} J Provine, Kelley Rivoire, 
\\Tomas Sarmiento, James Harris, and Jelena Vu\v{c}kovi\'{c}
\\{\small $^1$Department of Electrical Engineering, Stanford University, Stanford, California 94305 USA} }

\date{}
\maketitle

\begin{abstract}
We demonstrate a simple and rapid epoxy-based method for transferring photonic crystal cavities to the facets of optical fibers. Passive Si cavities were measured via fiber taper coupling as well as direct transmission from the fiber facet. Active quantum dot containing GaAs cavities showed photoluminescence that was collected both in free space and back through the original fiber. Cavities maintain a high quality factor (2000-4000) in both material systems. This new design architecture provides a practical mechanically stable platform for the integration of photonic crystal cavities with macroscale optics and opens the door for novel research on fiber-coupled cavity devices.  
\end{abstract}

\section{Main text}

Semiconductor photonic crystal (PC) cavities are micro-scale optical structures that possess interesting properties based on their ability to strongly localize light \cite{akahane, song}. To date, the vast majority of work on photonic crystal cavities has centered on the properties of devices still bound to their original growth substrates. This form factor is convenient for free space optical testing in the laboratory, but is difficult to integrate in larger systems consisting of many devices. Coupling light on- and off- chip is challenging due to severe size mismatch between PC components and external fiber optics. While certain applications of PCs such as dense all-optical processing may still require full on-chip integration, other applications of single or few photonic crystal cavities may benefit dramatically through a change in platform.        

In this letter, we report on an easy and rapid procedure to transfer PC cavities to fiber tips that avoids complicated microfab processing and uses ordinary epoxy as an adhesive layer. Our method can be done with a microscope-based setup in under an hour and requires only 10's of $\mu$m of precision for alignment. Additionally, material-specific recipes are not needed, and many types of active or passive cavities can be incorporated onto fibers. Using this technique, we functionalize optical fiber tips by successfully transferring Si cavities with resonances at 1500 nm as well as GaAs cavities resonant with InAs quantum dots (QDs) emitting at 1300 nm.

Photonic crystal cavities were fabricated through standard electron-beam lithography, dry etching, and undercutting. Both Si and GaAs air-bridged membranes were 220 nm thick and the GaAs material contained three layers of high density (300 dots/$\mu$m$^2$) InAs QDs with emission at 1300 nm. We use the common PC cavity design of a modified L3 defect with shifted air holes \cite{akahane} for high quality (Q) factor cavities and we use coupled cavity arrays (CCAs) for large mode area coupling \cite{hatice}. The L3 cavities in Si had a triangular lattice constant a = 450 nm and hole radius r = 0.22a while the cavities in GaAs had a = 330 nm and r = 0.22a. For the CCAs we use a = 490 nm and r = 0.38a and a two hole spacing between cavities in a square lattice \cite{hatice}. To facilitate alignment and guarantee that at least one L3 cavity will spatially overlap the fiber core (approximately 9 $\mu$m in diameter for SMF-28), we generate an array of uncoupled cavity devices with cavities spaced by no more than 9 $\mu$m apart. Similarly, we pattern a large 25 $\mu$m x 25 $\mu$m zone for the CCAs. A final outer region of 1 $\mu$m diameter air holes surrounding the cavities was used to release a larger 125 $\mu$m diameter circle for easier transfer.  

The first step in our transfer process is to use a sharp electrical probe to apply general purpose epoxy to the outer rim of a cleaved and stripped single-mode communication fiber facet (Fig. 1(a)). Epoxy is deposited in small amounts on the cladding surface so as to avoid contamination of the optical fiber core. A micromanipulator stage is used for positioning but we note that the precision required is low since the working fiber facet area is quite large. We next invert our fiber and approach our cavity structures from above. The fiber is centered over the cavities and then lowered until it comes into contact with the semiconductor membrane. Since the epoxy is applied on the cladding edges, it comes into contact with the outer release region of the membrane and does not spoil the central cavities. After the fiber is contacted with the PC cavities it can be withdrawn as the epoxy cures, carrying away the large membrane structure firmly bound to its facet. 
	
Fig. 1(b,c) show optical microscope pictures of a completed fiber plus PC cavity, or fiberPC, device for a silicon membrane with 1500 nm cavities. The slab is centered on the fiber facet and is affixed by a minimum amount of epoxy, seen in the image as the two darker areas. We find from angled SEM images that the Si membrane is almost perfectly planar to the facet surface with less than a micron in height variation at the edges where the epoxy is bonded. Our fiberPC device is robust and mechanically strong. Preliminary testing in various solution environments show no material degradation or membrane detachment. In addition, the functionalized fiber tip can be forcefully contacted with various hard surfaces without breaking. 

We first investigate the bound cavity properties of our Si L3 device with a fiber taper transmission measurement of a vertically oriented fiberPC (Fig. 2(a,b)) \cite{shambat}. Fig. 2(c) shows the transmission spectrum when the taper is placed on one of the central cavities of the fiberPC. From a Fano fit of the transmission dip at 1586 nm we find that the fundamental cavity mode loaded Q-factor is 2400 (Fig. 2(d)). Prior to transfer, these same cavities had measured fiber taper loaded Q-factors of 5000-10000. The decrease in Q-factor after transfer is most likely due to the lossy oxide cladding, as predicted by simulation. Therefore we find from these passive measurements that PC cavities can indeed survive relocation to fiber tips whilst maintaining a high Q. We also note that it is possible to use alternative oxide-clad cavity designs for ultra-high Q applications \cite{kuramochi}.

Next, we examine the ability to couple light between the fiber and cavities in a fiberPC using a direct transmission measurement (Fig. 3(a)). For this test we use fiberPC's with a CCA (Fig. 3(b)) since the large cavity modes overlap better with the Gaussian TE fiber modes, producing a stronger transmission resonance signal. Figures 3(c-e) show multiple transmission spectra for the fiberPC for several different collection points (taken by adjusting the area of collection with a pinhole). Clear cavity signals are obtained, indicating that is it possible to directly couple cavity resonances to the fiber. Additionally, the transmission lineshape is seen to be highly sensitive to the position of the pinhole, suggesting cavity radiation pattern dependence on the interference signal \cite{fan}.  

We turn our attention now to active GaAs fiberPC's with light-emitting quantum dots. Fig. 4(c) shows the PL spectrum of a single L3 cavity from a cavity array before transfer and still on-chip when pumped by 10's of $\mu$W of 830 nm light from a laser diode (LD). This same membrane was transferred onto a fiber tip and pumped through the fiber with the laser diode (Fig. 4(a)). As seen in Fig. 4(b), the QD PL is especially bright around the fiber core as well as around nearby cavities. When one of the cavities is spatially filtered with a pinhole from the full structure, we see a PL spectrum in Fig. 4(d) similar to that prior to transfer. A Lorentzian fit to the cavity mode gives a quality factor of 3700 (Fig. 4(e)), once again showing that strong resonances can be sustained at a fiber tip.

As a final measurement of a fiberPC device, we demonstrate PL excitation and collection in an `all-fiber' package. In this setup, the LD pump at 830 nm is sent to a 2 x 2 directional coupler before transmitting to the fiberPC (Fig. 5(a)). Rather than collecting the PL externally with bulky free space optics, the cavity PL that is reradiated into the fiber is collected in the return direction. Though the directional coupler is not optimized for 1300 nm, we still obtain clear spectra due to a strong PL signal (Fig. 5(b)). In this case the spectrum has a larger QD background component compared to the individual cavities because spatial filtering is not used and because the fiber core collects from a large area of uncoupled QD's. Two cavity fundamental modes are seen with slightly different wavelengths due to fabrication inhomogeneities. The cavities' alignment with the fiber core is likely to affect the intensity of collection, and additional studies to investigate this effect are underway. This measurement again proves that light can be coupled back and forth between PC cavities and a fiber in a monolithic package, far simpler and easy to use than a complete free space optical setup.

In summary, we have demonstrated a novel technique to functionalize optical fiber tips with semiconductor photonic crystal cavities. Our simple epoxy-based transfer process preserves robust cavity properties and can be applied toward numerous materials and cavity designs. The fiberPC platform enables the exploration of useful fiber-coupled PC devices and widely extends the range of possibilities for practical devices.

\section{Acknowledgements}
Gary Shambat and Kelley Rivoire were supported by the Stanford Graduate Fellowship. Gary Shambat also acknowledges the NSF GRFP for support. The authors acknowledge the financial support of the Center for Cancer Nanotechnology Excellence focused on Therapy Response (CCNE-TR) at Stanford University. Work was performed in part at the Stanford Nanofabrication Facility of NNIN supported by the National Science Foundation. We also acknowledge Bryan Park and Sonia Buckley for helpful discussions.

\section{List of captions}

FIG 1. \textbf{(a)} Fabrication flow for photonic crystal cavity transfer. (i) A minute amount of epoxy from a sharp metal tip is deposited on the face of a cleaved optical fiber. (ii) The fiber is inverted and brought into contact with the PC cavity array structure. (iii) Finally, the fiber is retracted from the semiconductor chip, carrying away cavities strongly held to its surface. \textbf{(b)} Angled optical microscope image of a completed fiberPC device. \textbf{(c)} Top-down microscope image of a fiberPC. The notches every 90 degrees are from tabs used to suspend the released membrane. The scale bar is 50 $\mu$m.    
\\
\\
FIG. 2. \textbf{(a)} Schematic of the fiber taper transmission measurement. Light from a broadband source (BBS) is sent through the fiber taper and the cavity transmission is recorded by an optical spectrum analyzer (OSA). OL is objective lens. \textbf{(b)} Optical image of the fiber taper touching a cavity for a vertically oriented fiberPC. The scale bar is 25 $\mu$m. \textbf{(c)} Transmission spectrum for the taper pictured in \textbf{(b)}. \textbf{(d)} Zoom-in of the fundamental cavity mode, revealing a loaded Q of 2400.
\\
\\
FIG. 3. \textbf{(a)} Schematic for the CCA transmission experiment. Light from the BBS is internally coupled to the fiberPC fiber and the transmitted signal is detected on the other side with a confocal microscope setup. \textbf{(b)} SEM image of a small subset of a larger CCA square. The scale bar is 1 $\mu$m. \textbf{(c-e)} Absolute transmission spectra of the CCA fiberPC for unique spot locations.
\\
\\
FIG. 4. \textbf{(a)} Schematic of the PL experiment. Quantum dots in GaAs cavities are pumped by a laser through the fiber while emission is collected with a free-space setup. \textbf{(b)} IR camera image of the QD PL when pumped internally. The fiber core is brightly seen in the center as well as nearby PC cavities. \textbf{(c)} PL spectrum of one cavity within the 5 x 5 array before PC membrane transfer. \textbf{(d)} PL spectrum of a different cavity within the original array now bound to the fiber showing similar results. \textbf{(e)} Close-up of the fundamental cavity mode from \textbf{(d)} along with a Lorentzian fit giving a Q of 3700.
\\
\\
FIG. 5. \textbf{(a)} Schematic of the `all-fiber' PL collection experiment. An 830 nm 2 x 2 directional coupler is placed in between the laser diode and fiberPC, with one return end sent to the spectrometer. \textbf{(b)} PL spectrum of the fiberPC emission that is collected back into the fiber, giving a similar response to the free-space measurement.

\newpage

\begin{figure}[htp]
\centering
\includegraphics{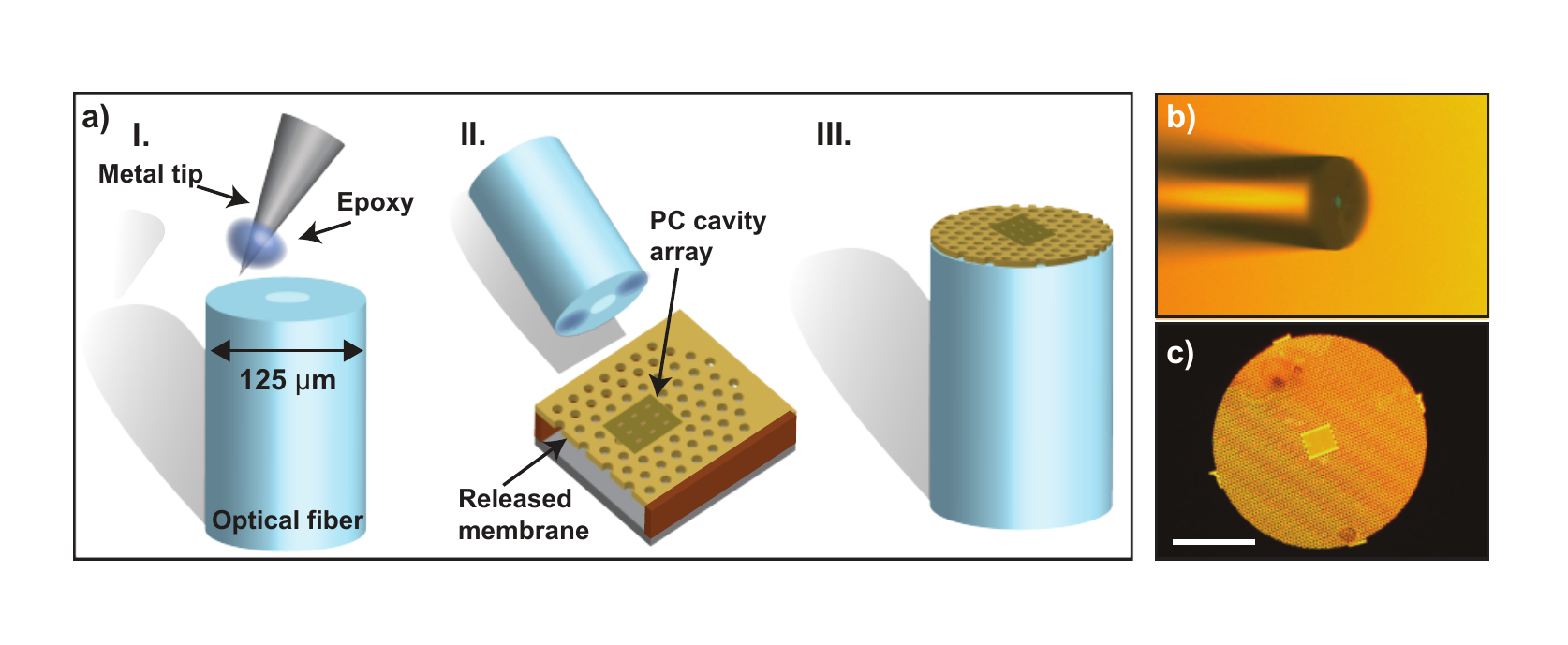}
\end{figure}

\newpage

\begin{figure}[htp]
\centering
\includegraphics{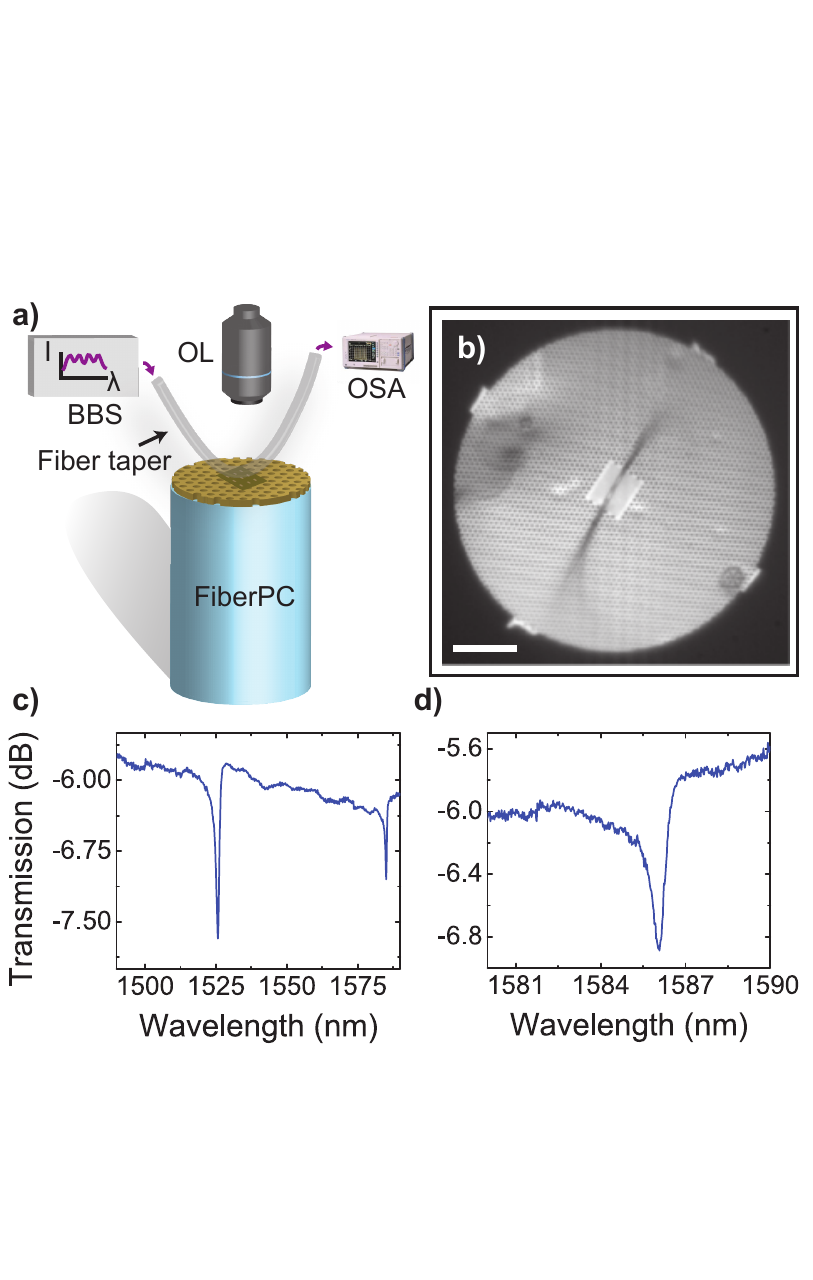}
\end{figure}

\newpage

\begin{figure}[htp]
\centering
\includegraphics{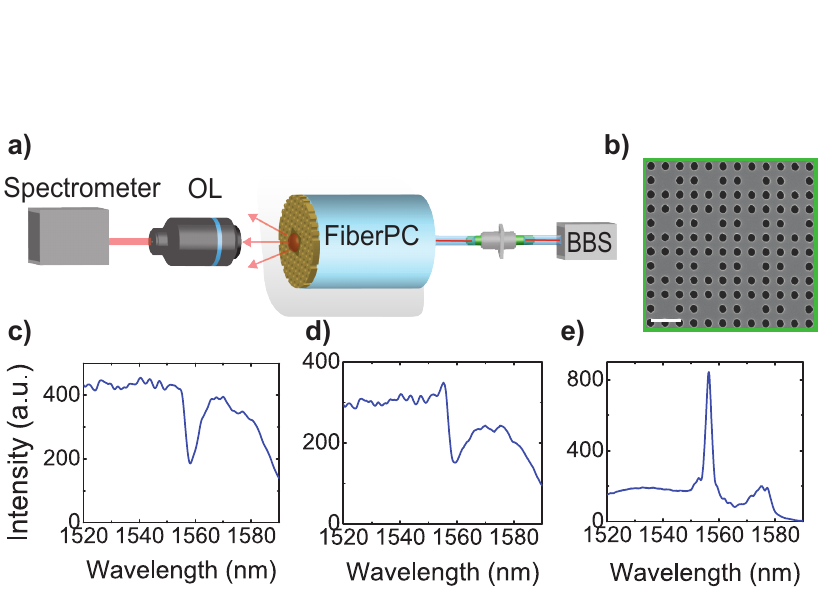}
\end{figure}

\newpage

\begin{figure}[htp]
\centering
\includegraphics{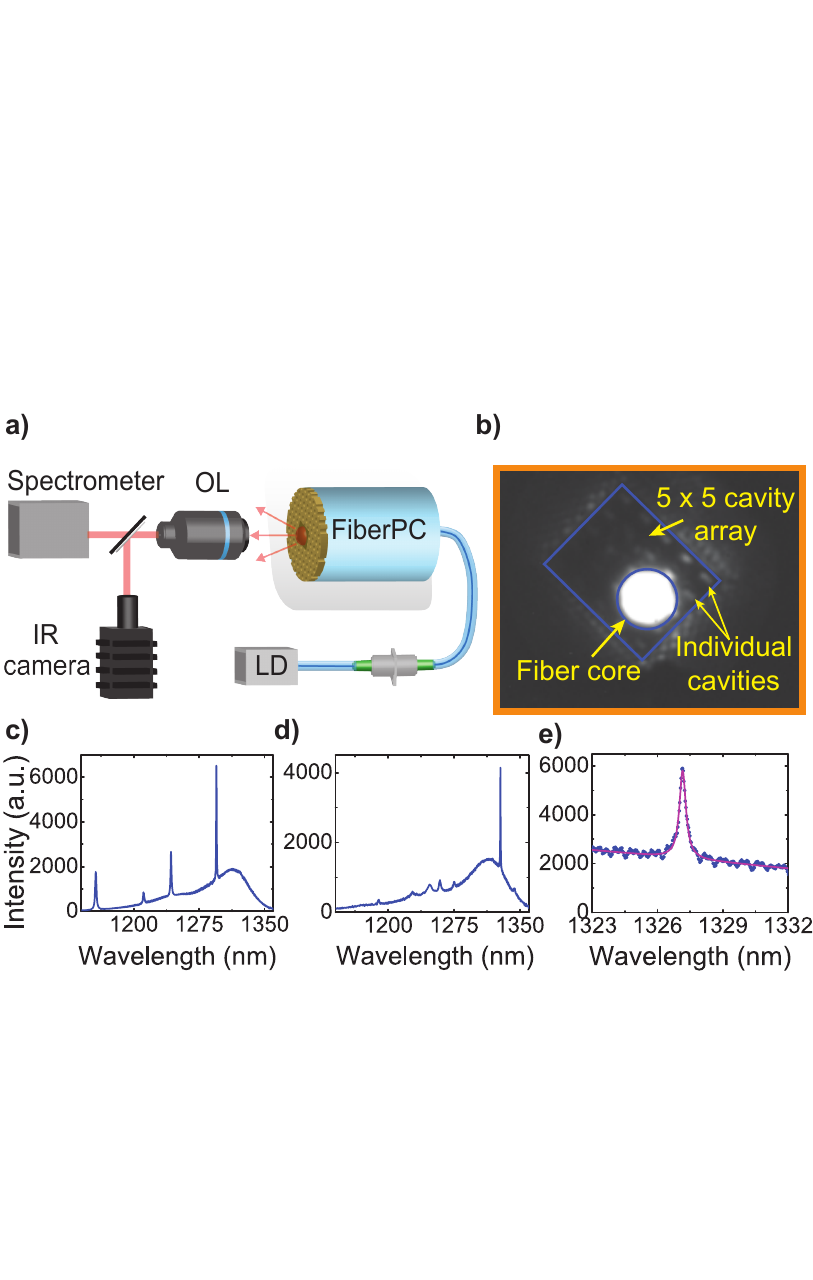}
\end{figure}

\newpage

\begin{figure}[htp]
\centering
\includegraphics{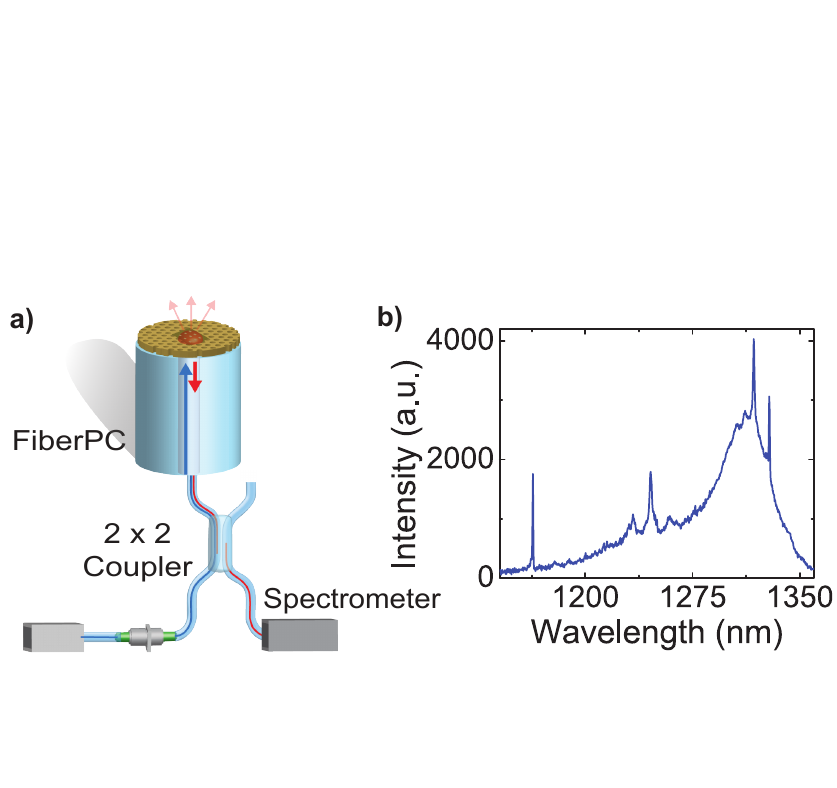}
\end{figure}

\end{document}